\documentclass[aps,prl,twocolumn,superscriptaddress]{revtex4-1}

\usepackage{amsmath}
\usepackage{amsthm}

\usepackage{bbm,graphicx,bm,hyperref}   
\usepackage[pdftex]{color}

\def\ave#1{\langle #1 \rangle}
\def\ii{{\rm i}}
\def\sx#1{\sigma^{\rm x}_{#1}}
\def\sy#1{\sigma^{\rm y}_{#1}}
\def\sz#1{\sigma^{\rm z}_{#1}}
\def\tx#1{\tau^{\rm x}_{#1}}
\def\ty#1{\tau^{\rm y}_{#1}}
\def\tz#1{\tau^{\rm z}_{#1}}
\def\Z2{\mathbbm{Z}_2}

\def\tr#1{{\rm tr}(#1)}
\def\1{\mathbbm{1}}

\def\ket#1{{| #1 \rangle}}

\def\tit#1{{\em #1},}

\usepackage{booktabs}
\usepackage{multirow}
\AtBeginDocument{
\heavyrulewidth=.08em
\lightrulewidth=.05em
\cmidrulewidth=.03em
\belowrulesep=.65ex
\belowbottomsep=0pt
\aboverulesep=.4ex
\abovetopsep=0pt
\cmidrulesep=\doublerulesep
\cmidrulekern=.5em
\defaultaddspace=.5em
}

\begin{document}

\title{Exact localized and ballistic eigenstates in disordered chaotic spin ladders\\
and the Fermi-Hubbard model}

\author{Thomas Iadecola}
\affiliation{Joint Quantum Institute and Condensed Matter Theory Center, Department of Physics, University of Maryland, College Park, Maryland 20742, USA}
\author{Marko \v Znidari\v c}
\affiliation{Physics Department, Faculty of Mathematics and Physics, University of Ljubljana, 1000 Ljubljana, Slovenia}
\affiliation{Abdus Salam ICTP, Strada Costiera 11, 34151 Trieste, Italy}

\date{\today}

\begin{abstract}
We demonstrate the existence of exact atypical many-body eigenstates in a class of disordered, interacting one-dimensional quantum systems that includes the Fermi-Hubbard model as a special case.  These atypical eigenstates, which generically have finite energy density and are exponentially many in number, are populated by noninteracting excitations. They can exhibit Anderson localization with area-law eigenstate entanglement or, surprisingly, ballistic transport at any disorder strength. These properties differ strikingly from those of typical eigenstates nearby in energy, which we show give rise to diffusive transport as expected in a chaotic quantum system. We discuss how to observe these atypical eigenstates in cold-atom experiments realizing the Fermi-Hubbard model, and comment on the robustness of their properties.
\end{abstract}



\maketitle

{\em \bf Introduction.--} 
The standard theory of matter uses equilibrium statistical ensembles to classify all possible phases according to local order parameters. This classical picture was shattered in recent decades by two important realizations. First, there are topological phases that are indistinguishable by local order parameters. Second, there are {\em eigenstate phases}~\cite{eigenstaterew} that arise when a system fails to thermalize~\cite{thermalrew} and standard equilibrium ensembles do not describe the late-time dynamics. Because eigenstate phases are athermal, standard no-go theorems prohibiting equilibrium phase transitions in, e.g., one-dimensional systems do not apply, opening a new world of possibilities.

While the general conditions for the occurrence of such phases are not known, disordered systems provide a paradigmatic example: for sufficiently strong disorder a many-body localized (MBL) phase can appear~\cite{MBLrew,mbl2,alet18,mbl3}. One exciting feature of such localized systems is that they can preserve quantum order at infinite temperature~\cite{huse13,nayak13,anushya14}, enabling, e.g., the storage of quantum information~\cite{moore15,edge}. One therefore has an interesting interplay of interactions, disorder, and symmetry.

A natural question is whether there is new interesting physics between the two extremes represented by thermalizing and nonergodic systems. The answer is yes. In clean systems a so-called quantum disentangled liquid has been proposed~\cite{Grover14,garrison,essler17} where some degrees of freedom have an area-law entanglement entropy. 
Related but different possibilities include weak ergodicity breaking in clean one-dimensional (1D) systems due to local Hilbert-space constraints~\cite{turner}, dynamical bottlenecks~\cite{mauro,levi15,juan,alex}, energy-scale separation~\cite{schiro12,moore16}, or an effective initial-state disorder due to conserved quantities~\cite{smith1,smith2,smith3} or gauge invariance~\cite{marcello}. While in 1D spin-1/2 systems sufficiently strong disorder will cause full MBL, in higher dimensions~\cite{2dexper} one might expect delocalization~\cite{deroeck} (see though~\cite{anto16}) and ergodicity. An intermediate regime
where rich new phases might be possible 
is disordered spin ladder models or, equivalently, systems with 
onsite Hilbert space dimension greater than two.
Such models arise naturally in systems with symmetries, such as experimental implementations~\cite{schreiber15,1dcoupled} of the disordered Fermi-Hubbard chain~\cite{rigol15}, which have an $SU(2)$ spin-rotation symmetry. One can argue that non-Abelian symmetries favor delocalization due to the presence of highly degenerate multiplets~\cite{potter16,dima17,anushya14}. This is indeed what happens: spin and charge degrees of freedom behave markedly differently~\cite{prb16}, and there is an ongoing discussion~\cite{sarang17,sirker17,bonca17,marcin18,kuba,peter18,clark18,dima18} on the ultimate fate of such $SU(2)$ symmetric systems, and more generally of models with enlarged onsite Hilbert spaces and discrete non-Abelian~\cite{sid,prakash} or Abelian~\cite{iadecola18} symmetries.

We study the role of symmetries in a class of interacting systems with onsite disorder. By an explicit construction we prove the existence of exponentially large invariant subspaces that are either ballistic or localized regardless of the disorder or interaction strength, and are present in integrable as well as in chaotic models, irrespective of $SU(2)$ symmetry. This shows that even an innocuous looking system, classified as quantum chaotic according to eigenlevel statistics, can violate the strong eigenstate thermalization hypothesis~\cite{biroli10,alessio16,deutsch18}, stating that in the thermodynamic limit all eigenstates should be thermal. Furthermore, the class of models discussed includes the disordered Fermi-Hubbard chain realized in recent experiments~\cite{schreiber15} probing MBL; we thus rigorously show the importance of symmetries and that choosing specific simple initial states can fundamentally influence dynamics in such experiments.

{\em \bf Models.--}
We study a class of two-leg spin ladders with the Hamiltonian
\begin{subequations}
\label{eq:XXladder}
\begin{align}
\label{eq:XXladdera}
H&=H^{||}+H^{\perp}=\frac{1}{4}\sum_{k=1}^{L-1} h^{||}_{k,k+1}+\frac{1}{4}\sum_{k=1}^{L} h^{\perp}_{k},
\end{align}
where
\begin{align}
\label{eq:XXladderb}
h^{||}_{k,k+1}&=\sx{k} \sx{k+1}+\sy{k}\sy{k+1}+\tx{k}\tx{k+1}+\ty{k}\ty{k+1},\\
h^{\perp}_{k}&=J(\sx{k} \tx{k}+\sy{k} \ty{k})+\Delta_k\, \sz{k} \tz{k}+h_k(\sz{k}+\tz{k}).\nonumber
\end{align}
\end{subequations}
Rungs of the ladder are labeled by $k=1,\dots,L$, and spins on the upper and lower legs of the ladder are represented by Pauli matrices $\sigma^\alpha_{k}$ and $\tau^\alpha_{k}$ ($\alpha=\rm x,\rm y,\rm z$), respectively. In numerical examples $h_k$ is drawn uniformly at random from the interval $[-h,h]$. For $J=0$ the model is equivalent to the Fermi-Hubbard model by a Jordan-Wigner transformation~\cite{Shastry86} (spins on the upper and lower legs correspond to spin-up and -down fermions, respectively, and $\Delta_k$ corresponds to the onsite Hubbard interaction).
\begin{figure}[t!]
\centerline{\includegraphics[width=.8\columnwidth]{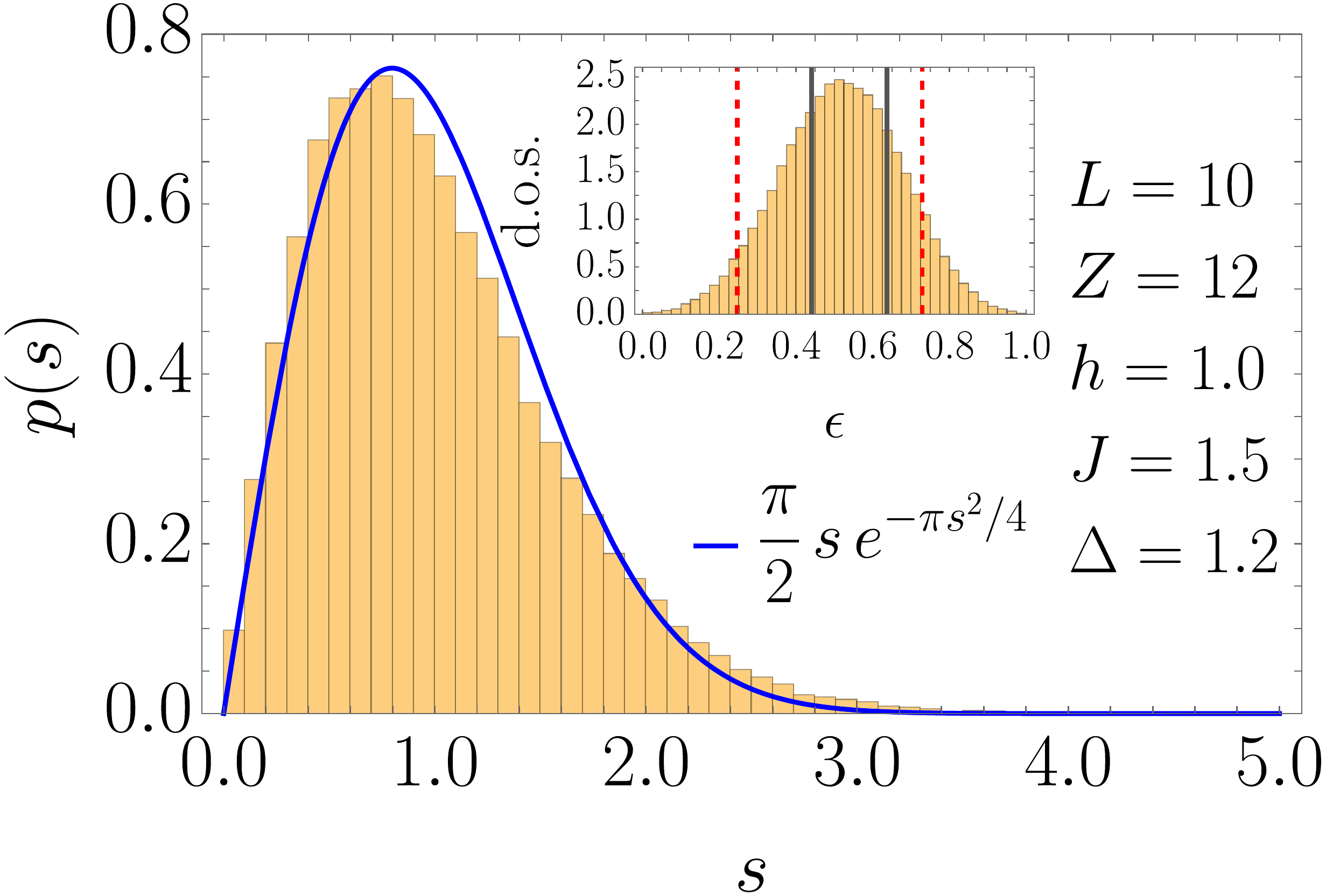}}
\caption{Level spacing distribution for a generic instance of the model~\eqref{eq:XXladder}, averaged over 100 disorder realizations.  The distribution is well-described by the Wigner-Dyson distribution (blue).  Inset: Disorder-averaged density of states vs.~energy density $\epsilon$ (normalized to lie within the interval $[0,1]$) for the same model.  Eigenstates corresponding to energies between the black vertical lines were used to accumulate the statistics in the main figure.  Dashed red vertical lines denote the bandwidth of a six-doublon invariant subspace, demonstrating the finite energy density of eigenstates within this subspace.}
\label{fig:LSD}
\end{figure}

The model \eqref{eq:XXladder} has a $U(1)$ symmetry associated with the total magnetization $Z=\sum^L_{k=1}(\sz{k}+\tz{k})$. It also has a $\Z2$ symmetry $\sigma^\alpha_{k}\leftrightarrow\tau^\alpha_{k}$. In the Hubbard case ($J=0$) one has an additional
$SU(2)$ spin-rotation symmetry~\cite{1dHubbard}. This motivates the definition of ``charge" and ``spin" densities $d_k \equiv \frac{1}{2}(\sz{k}+\tz{k})$ and $s_k \equiv \frac{1}{2}(\sz{k}-\tz{k})$, respectively.  In the Hubbard language, $\langle d_k\rangle=\pm1$ corresponds to the presence of a doublon/holon, while $\langle s_k\rangle=\pm1$ corresponds to the presence of a spin-up/down fermion.

{\em \bf Quantum ergodicity.--}
We now demonstrate that the class of Hamiltonians in Eq.~\eqref{eq:XXladder} is generically quantum chaotic when the disorder strength $h$ is not too large (the clean limit of Eq.~\eqref{eq:XXladder} was studied in Ref.~\cite{PRL13}). This will establish the atypicality of the special eigenstates constructed below.  One common indicator of quantum chaos is the distribution of the spacing $s$ between adjacent many-body energy levels~\cite{Haake}. After averaging over disorder, once all symmetries have been resolved and energies corresponding to the atypical eigenstates have been removed, we find a distribution consistent with Wigner-Dyson statistics typical of chaotic systems, see Fig.~\ref{fig:LSD}.  

\begin{figure}[t!]
\centerline{\includegraphics[width=3.1in]{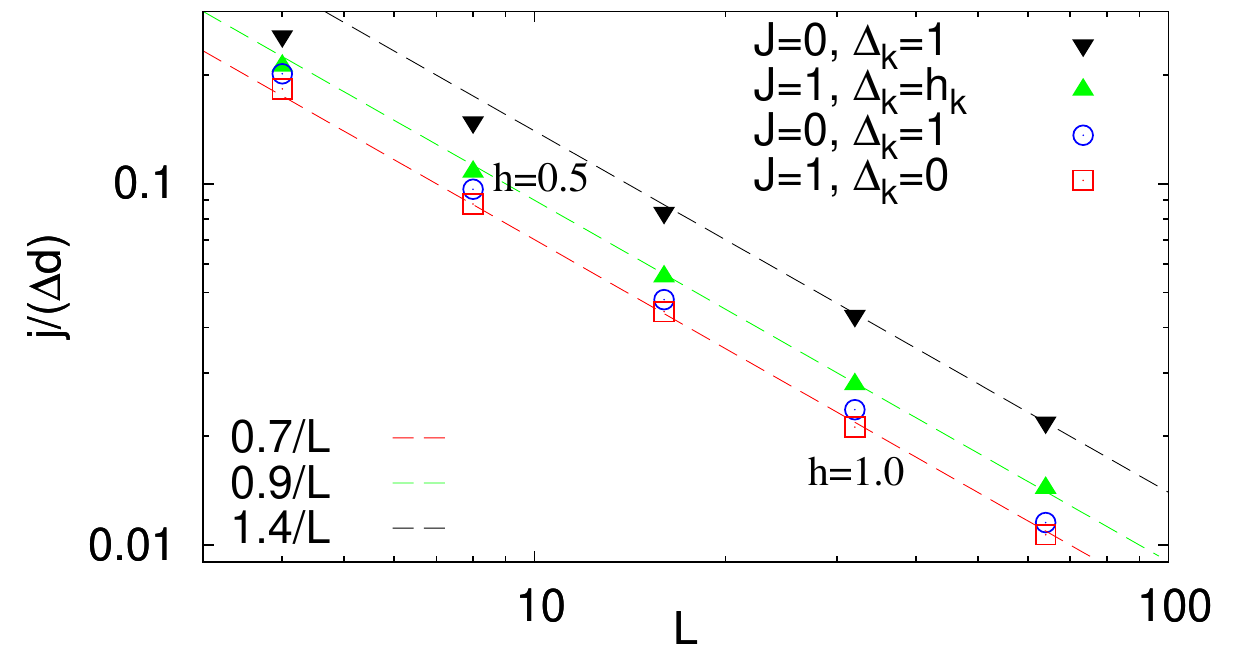}}
\caption{Diffusive NESS charge transport for weakly disordered models in the full Hilbert space (full triangles for disorder $h=0.5$, empty symbols for $h=1$). Both the Hubbard case (down triangles and circles) and a generic ladder (squares and triangles) are shown.}
\label{fig:Lindblad}
\end{figure}

Another indicator of quantum ergodicity is diffusive transport, which we demonstrate arises in the model \eqref{eq:XXladder} when disorder is sufficiently weak. Focusing on high temperature (energy density) transport in large systems we employ a boundary driven Lindblad master equation~\cite{Lindblad,Lindblad2}. We use four Lindblad operators that raise/lower the magnetization at site $k=1$ (two for each ladder leg), and four that act at $k=L$. Details about the method and driving, which induces transport of the charge $d_k$ (which is conserved also for $J \neq 0$), can be found in e.g.~\cite{PRB12}, where a clean Hubbard model was studied. At late times the solution 
$\rho(t)$ of the Lindblad master equation reaches a unique nonequilibrium steady state (NESS) $\rho_\infty$. Transport is probed by calculating the $L$-dependence of the NESS current expectation value, $j\equiv\tr{\rho_\infty j^{(d)}_k}$, with $j^{(d)}_k\equiv\frac{1}{4}(\sx{k}\sy{k+1}-\sy{k}\sx{k+1})+\frac{1}{4}(\tx{k}\ty{k+1}-\ty{k}\tx{k+1})$ that satisfies $\dot{d}_k=j^{(d)}_k-j^{(d)}_{k-1}$. For large $L$ the boundary density is $\tr{\rho_\infty d_{1,L}}\approx \pm \mu$ and so for our weak driving $\mu=0.1$ we have $\Delta d\equiv\ave{d_1}-\ave{d_L}\approx 0.2$. For small disorder $h=0.5$ or $h=1$ the charge density profiles (not shown) are linear, as expected for diffusion. In Fig.~\ref{fig:Lindblad} we show that $j \sim 1/L$, so that the diffusive law $j = D \frac{\Delta d}{L}$ holds. We stress that the observed diffusion is not a consequence of boundary driving but a true property of the bulk~\cite{nesskubo18}.

{\em \bf Invariant subspaces.--}
We now explicitly construct a number of invariant subspaces of the Hamiltonian \eqref{eq:XXladder} that will enable us to analytically show the existence of ballistic and/or localized eigenstates irrespective of the values of the parameters $J$, $h_k$, and $\Delta_k$. Let us take for a local basis the eigenstates of $h_k^\perp$. Denoting by $0$ and $1$ the two eigenstates of $\sz{}$ with eigenvalues $+1$ and $-1$, respectively, and by, e.g., $\ket{{1 \atop 0}}$ a rung state with $1$ in the upper leg and $0$ in the lower, we have the four rung eigenstates described in Table~\ref{tab:dict}. Products of rung eigenstates, i.e., $\ket{\alpha_1\ldots \alpha_L}$, $\alpha_k \in \{S,T,H,D\}$, are manifestly eigenstates of the total rung Hamiltonian $H^\perp$ with eigenenergies $E_\perp^{\bm{\alpha}}=\frac{1}{4}\sum_k E^{\alpha_k}_k$. Following Ref.~\cite{PRL13}, the leg Hamiltonian has a very simple action on some basis states. Namely, acting on two neighboring rungs gives
\begin{align}
\begin{split}
  h^{||}_{k,k+1}\ket{\alpha_k \beta_{k+1}}&=2\ket{\beta_k \alpha_{k+1}},\qquad \beta_j \in \{ S,T\},  \\
  h^{||}_{k,k+1}\ket{\beta_k \alpha_{k+1}}&=2\ket{\alpha_k \beta_{k+1}},\qquad \alpha_j \in \{H,D\},
\end{split}
  \label{eq:hop}
\end{align}
i.e., if a doublon or a holon meets a singlet or a triplet they just exchange positions. Eq.~(\ref{eq:hop}) specifies $H^{||}$'s action on $8$ of the $16$ two-rung basis states. Four more important relations are the annihilations
\begin{equation}
  h^{||}_{k,k+1}\ket{ \{ ST,TS,HH,DD\} }=0.
  \label{eq:zero}
\end{equation}
The action on the remaining four two-rung basis states~\cite{foot1} induces a nontrivial dynamics outside of the invariant subspaces, and will not be needed here. 

Using relations \eqref{eq:hop} and \eqref{eq:zero} we can readily construct invariant subspaces. First, we observe that the states $\ket{STST\cdots}$ and $\ket{TSTS\cdots}$ are annihilated by $H^{||}$ -- 
they are ``vacuum'' states (inert backgrounds). If we now insert into one of these vacuum states an arbitrary number of only doublons, or only holons, such a subspace will be invariant under $H^{||}$. Starting with, e.g., two holons $\ket{STS{H}_jTS{H}_kTS\cdots}$, a repeated action of $H^{||}$ will only move the two holons around to all possible $L \choose 2$ positions $j,k$, preserving the number of each of the four letters. Similar behavior arises upon inserting only doublons. Inserting $r$ doublons (or holons) results in an $L \choose r$-dimensional invariant subspace of $H$ (\ref{eq:XXladder}). The total dimension of all such invariant subspaces is $2^{L+2}$.

\begingroup
\squeezetable
\begin{table}[t!]
\begin{ruledtabular}
\begin{tabular}{cccc}
Eigenstate & Notation & $\ave{d_k}$ &Eigenenergy \\
\midrule
singlet & $\ket{S}:=\frac{1}{\sqrt{2}}\left(\ket{{0 \atop 1}}-\ket{{1 \atop 0}}\right)$ & $0$ & $E^S_k=-2J-\Delta_k$\\
triplet & $\ket{T}:=\frac{1}{\sqrt{2}}\left(\ket{{0 \atop 1}}+\ket{{1 \atop 0}}\right)$ & $0$ & $E^T_k=2J-\Delta_k$\\
doublon & $\ket{D}:=\ket{{0 \atop 0}}$ & $+1$ & $E^D_k=\Delta_k+2h_k$\\
holon & $\ket{H}:=\ket{{1 \atop 1}}$ & $-1$ & $E^H_k=\Delta_k-2h_k$
\end{tabular}
\end{ruledtabular}
\caption{Notation for eigenstates of $h_k^\perp$ [see (\ref{eq:XXladderb})], with corresponding eigenenergies and charge densities $d_k = \frac{1}{2}(\sz{k}+\tz{k})$.}
\label{tab:dict}
\end{table}
\endgroup

The Hamiltonian describing the dynamics within an invariant subspace has constant off-diagonal elements for each possible hop of $D$ or $H$, and diagonal elements given by the total rung eigenenergies $E^{\bm \alpha}_\perp$. The dynamics is that of {\em noninteracting particles}, i.e., a tight-binding model with onsite energies given by the eigenenergies of the local rung states $\alpha_k$ (Table~\ref{tab:dict}). Depending on the choice of $\Delta_k$ and $h_k$, we can have rich free physics embedded within a seemingly generic (chaotic or integrable) model. Let us single out some of the more interesting examples: (i) constant $\Delta_j\equiv\Delta$ and uniformly random $h_j$ results in Anderson localization in any doublon or holon subspace for any disorder strength (their onsite disorder is simply $\pm 2h_j$, see Table \ref{tab:dict}); (ii) disordered $\Delta_j=h_j$ causes a constant energy offset $-h_k$ for all four basis states while the doublon state has an additional onsit
 e energy of $+4h_j$, leading to Anderson localization in the doublon subspaces and ballistic transport in the holon subspaces, again irrespective of the disorder strength (see~\cite{barlev} for a study of an interaction-disordered Hubbard model); (iii) by using either case (i) or case (ii) with a quasiperiodic potential $h_j=\lambda \cos{(2\pi \kappa j+\phi)}$ one can realize the Aubry-Andr\' e-Harper model~\cite{AAH,AAH2}, which features ballistic, diffusive, or localized dynamics depending on $\lambda$.

The subspaces constructed here do not seem to be related to any local conserved quantity. They exist irrespective of the presence or absence of $SU(2)$ symmetry or integrability (they arise both in the integrable clean Hubbard case and the chaotic case with $J\neq0$). Crucial for their existence is the simple nature of $H^{||}$ and the presence of a $\Z2$ symmetry leading to a decoupling of the singlet and triplet states from disorder.

\begin{figure}[t!]
\centering
\includegraphics[width=.95\columnwidth]{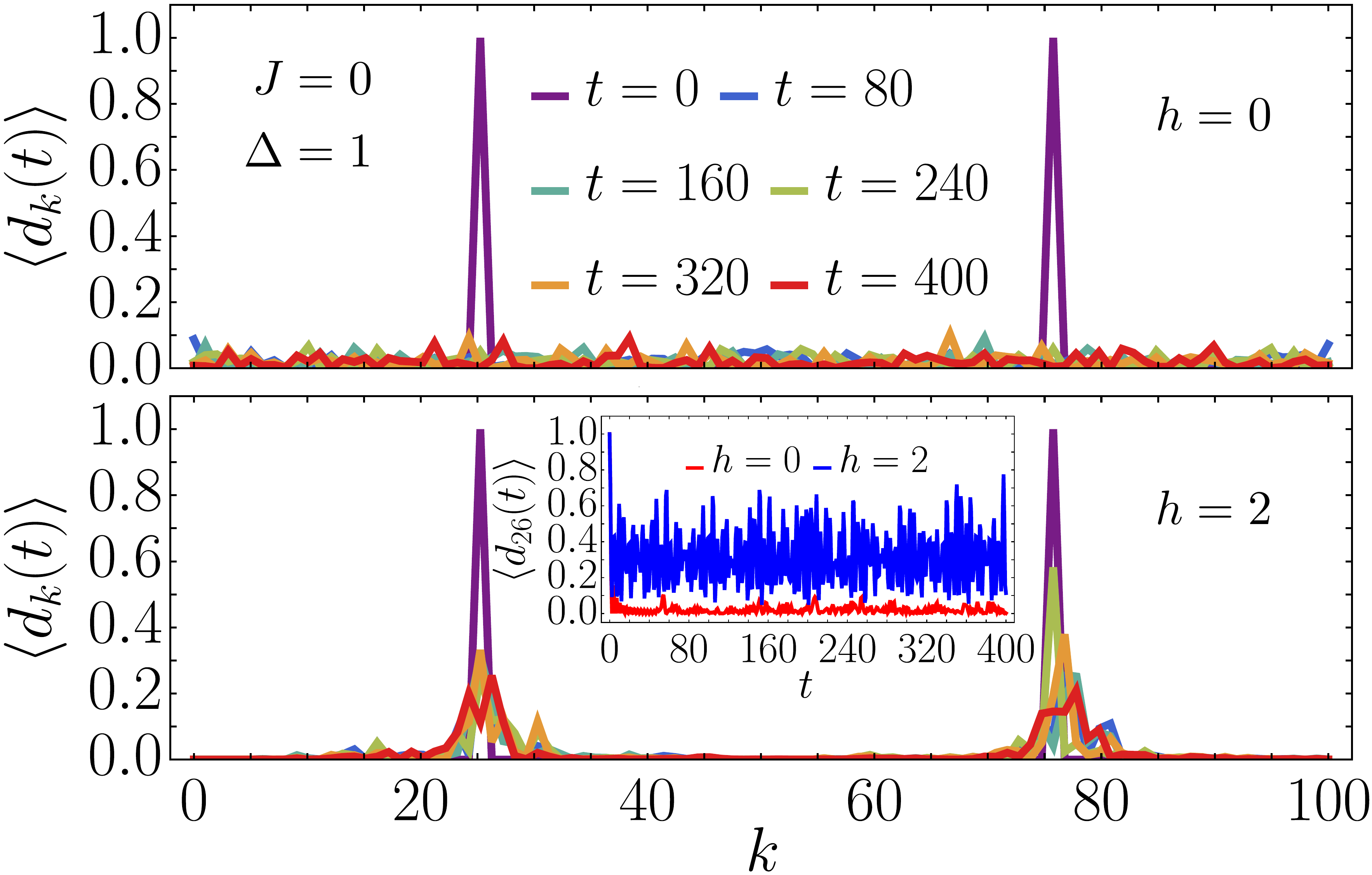}
\caption{Ballistic (top) vs. localized (bottom) charge dynamics in a two-doublon invariant subspace in the Fermi-Hubbard model from exact diagonalization at $L=100$. The charge density profile is shown for various times $t=0,\dots,400$. Inset: Charge density vs.~$t$ on a single site initially populated by a doublon for $h=0$ (red) and $h=2$ (blue). The dynamics outside of the invariant subspaces is ergodic for $h=2$.}
\label{fig:evol}
\end{figure}
We emphasize that invariant subspaces with a finite density of doublons or holons are exponentially large in $L$ and yield finite-energy-density eigenstates of the Hamiltonian \eqref{eq:XXladder} because the bandwidth of such states scales with the doublon/holon number (see inset of Fig.~\ref{fig:LSD}).  Such eigenstates are thus highly atypical: Anderson-localized subspaces [cases (i)--(iii)] host eigenstates with area-law entanglement that do not contribute to transport, while subspaces that decouple from the disorder [cases (ii)--(iii)] host volume-law-entangled eigenstates that provide a ballistic contribution to transport.  In contrast, generic eigenstates of the model \eqref{eq:XXladder} both exhibit volume-law entanglement and contribute to diffusive transport. Moreover, these invariant subspaces are spanned by simple product states that are experimentally accessible, as we discuss below. 

{\em \bf Experimental implementation.--}
The very fact that we have localized subspaces should facilitate experimental observation: clear ergodicity breaking can be observed by preparing an initial product state in such a subspace. This is illustrated in Fig.~\ref{fig:evol} for the important special case of the Hubbard model. Starting from an initial state containing two doublons in an $ST$ background, the initial charge stays localized for $h=2$, while it relaxes ballistically in the clean Hubbard model ($h=0$).  Even if the initial state is prepared imprecisely and does not lie exclusively within a localized subspace, its time evolution will show a localized component as long as it has finite overlap with one such subspace.

All ingredients needed for such an evolution have been experimentally realized. The initial states can be chosen to be simple product states. However, they are not in the natural experimental basis of local-density product states with definite fermion spin in the $z$-basis.
To reach these states, two crucial ingredients are required.  First, one needs the ability to prepare initial states with a controlled charge density and spin profile; progress in this direction has recently been made with fermionic quantum gas microscopes~\cite{microscope1,microscope2,microscope3}. Moreover, high-fidelity single-site addressing of the spin state in a bosonic optical lattice was demonstrated in~\cite{blochSingleSpin}.  An example of a relevant initial state to prepare to reach the subspace containing one doublon on site $k_0$ in a $TS$ background is
\begin{align}
\label{eq:state}
\dots
c^\dagger_{\downarrow,k_0-2}c^\dagger_{\uparrow,k_0-1}
(c^\dagger_{\downarrow,k_0}\!c^\dagger_{\uparrow,k_0})
c^\dagger_{\downarrow,k_0+1}
c^\dagger_{\uparrow,k_0+2}\dots
\ket{\Omega},
\end{align}
where $c^\dagger_{\uparrow/\downarrow,k}$ creates a spin-up/down fermion on site $k$ and $\ket{\Omega}$ is the vacuum. Performing a global spin rotation $U=\exp{(-\ii \frac{\pi}{2} S_{\rm y})}$, where $S_{\rm y}=\frac{\ii}{2}\sum_k c^\dagger_{\downarrow,k}c_{\uparrow,k}-c^\dagger_{\uparrow,k}c_{\downarrow,k}$, then brings the $z$-basis fermion spin states into the $x$-basis~\cite{foot3}. In practice such a rotation could be performed in two steps, rotating first around the $x$-axis (by resonant pumping between hyperfine levels) and then the $z$-axis (using a Zeeman shift).  The necessary ingredients to perform such rotations have already been demonstrated in experiments~\cite{blochSingleSpin,spielmanSOC}, even with single-site resolution~\cite{blochSingleSpin}. 

\begin{figure}[t!]
\vskip5pt
\centerline{$\log_{10}|\ave{d_k}|$ \hskip1.8cm $\log_{10}|\ave{s_k}|$}
\vskip-5pt
\centerline{\includegraphics[height=3.92cm]{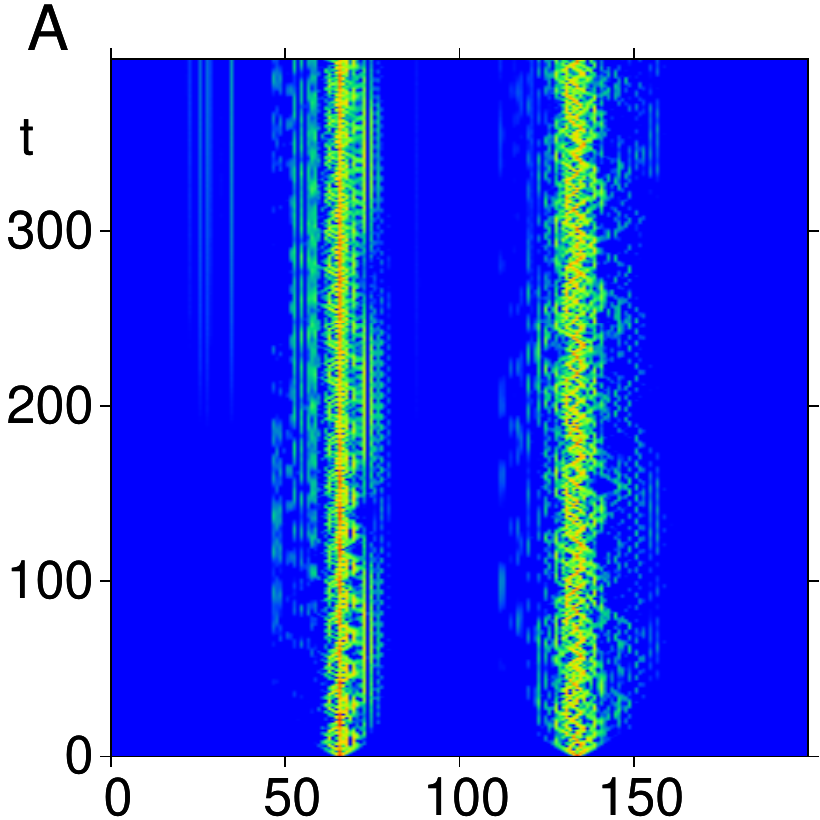}\includegraphics[height=3.92cm]{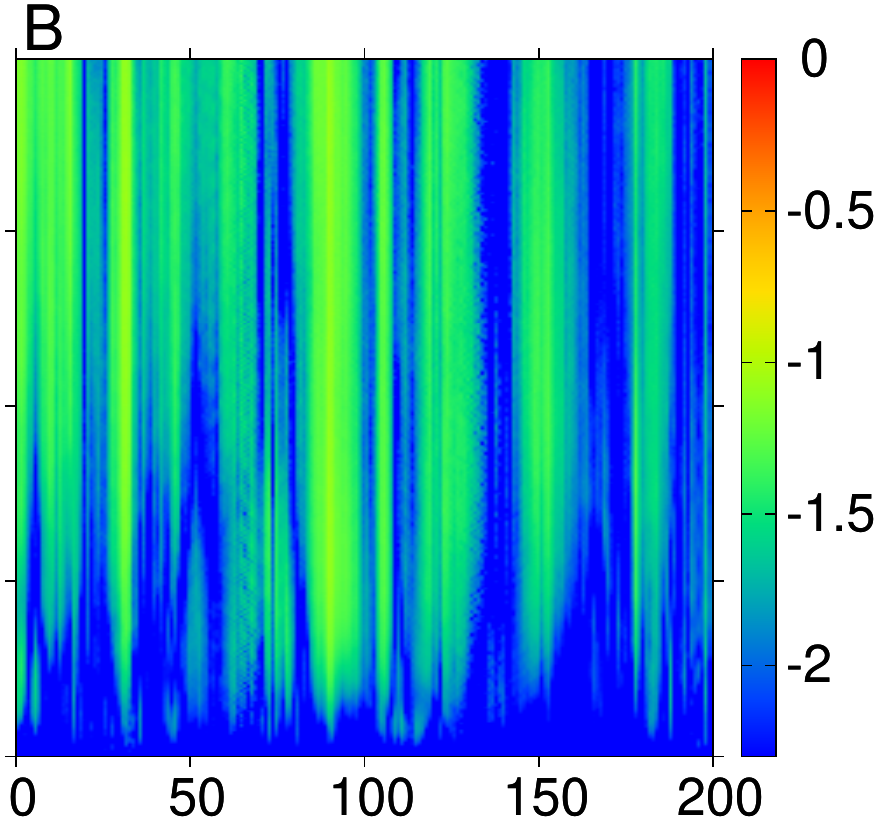}}
\vskip-1pt
\centerline{\hskip3pt\includegraphics[height=4.2cm]{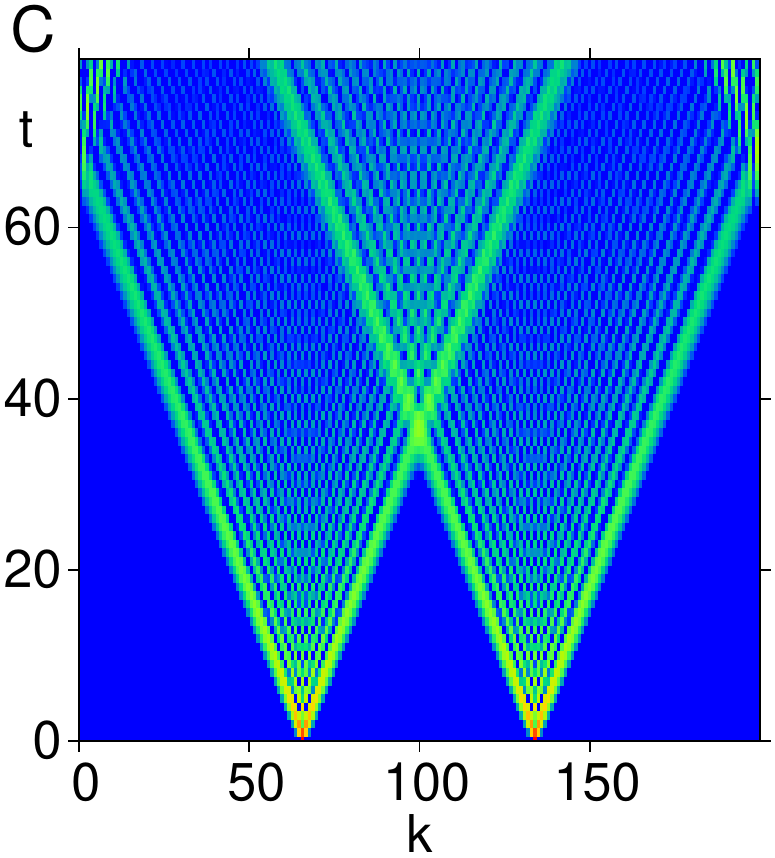}\includegraphics[height=4.2cm]{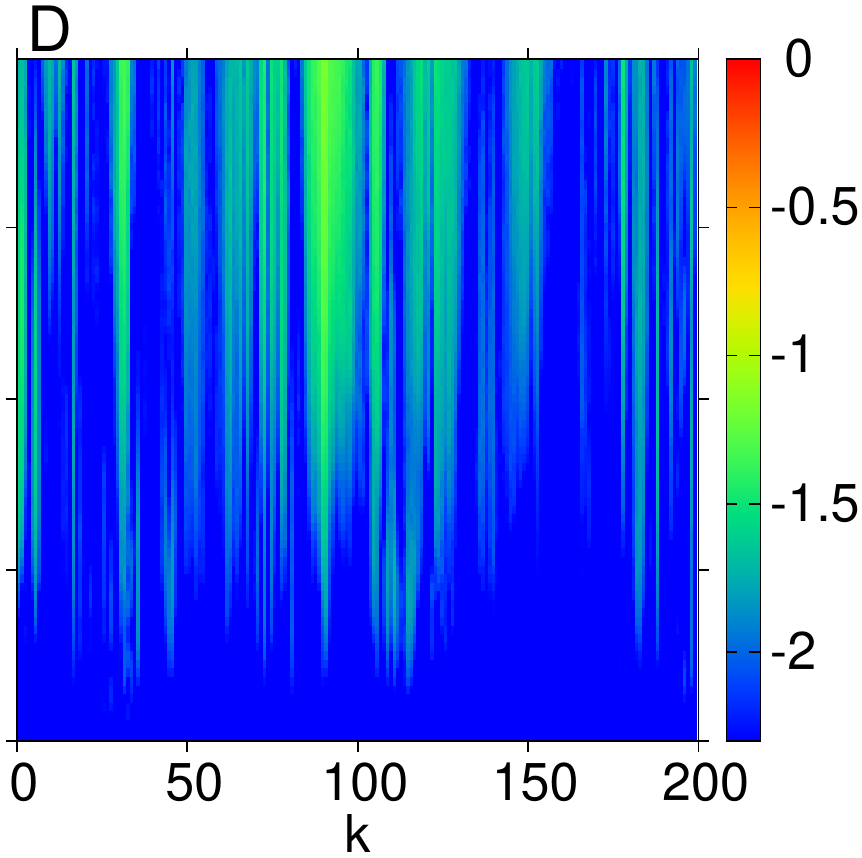}}
\caption{Evolution for a two-doublon (A and B), and a two-holon initial state (C and D) for weak $4\%$ breaking of $\Z2$ symmetry ($\varepsilon=0.04$). All data are for $J=0$, $\Delta_j=h_j$, $L=200$, and the same disorder realization in all frames. Color (shown in a log-scale) shows the charge (A and C) and spin density (B and D).}
\label{fig:n200}
\end{figure}

In practice one also must contend with an inexact realization of the model, including, e.g., weak breaking of the $\Z2$ leg-exchange symmetry (see, e.g., \cite{pranjal16}). We numerically check how much such weak symmetry breaking changes the dynamics on experimentally relevant timescales. To that end we replace the $\Z2$-symmetric disorder in \eqref{eq:XXladder} with slightly asymmetric onsite fields, $h_k \sz{k}+h_k' \tz{k}$, where $(h_k'-h_k)/h_k=\xi_k$ is a random number uniformly distributed over $[-\frac{\varepsilon}{2},\frac{\varepsilon}{2}]$. Using a matrix product state ansatz we evolve the initial state under this modified Hamiltonian with $\varepsilon=0.04$, $J=0$ and $\Delta_k=h_k$, starting with either a two-holon or a two-doublon initial state. Recall that in this case, for perfect $\Z2$ symmetry, the doublons are localized, while the holons are ballistic. Results are shown in Fig.~\ref{fig:n200}.
We can see that some spin density $s_k(t)$ is indeed created on a timescale $\sim 1/\varepsilon$ (for $\varepsilon=0$ one has $s_k(t)\equiv 0$), but grows only to about $10^{-1}$ with time. The charge density $d_k(t)$ on the other hand shows the same localized/ballistic dynamics as without the symmetry breaking. We note that for the Hubbard model with $\Delta_k\equiv\Delta$, where holons and doublons are both localized, the dynamics for $\varepsilon \neq 0$ (data not shown) is essentially the same as in Figs.~\ref{fig:n200}A and B. Thus the charge dynamics is rather resilient for weak symmetry breaking.

{\em \bf Conclusion.--}
We have explicitly constructed a new class of exponentially many ballistic or localized eigenstates embedded in an otherwise (in general) chaotic model. The phenomenon is exact and independent of interaction or disorder strength, in contrast to, e.g., localized states observed numerically for strong disorder in ladder systems in Refs.~\cite{clark18,iadecola18}. While our construction includes the disordered $SU(2)$-symmetric Fermi-Hubbard model as an important special case, the result is more general and does not rely on the presence of $SU(2)$ symmetry. In this sense it is also different from the (polynomially many) special exact eigenstates with $\sim \log{L}$ entanglement entropy identified in non-integrable $SU(2)$-symmetric models~\cite{yang,bernevig1,clark18,bernevig2}.

Our results also shed light on the question of localization in systems with non-Abelian symmetries. While it is known that disorder can protect spontaneous symmetry breaking~\cite{huse13,nayak13,anushya14}, we find, intriguingly, that a reverse effect is possible---a global $\Z2$ symmetry can protect exponentially many localized eigenstates against delocalization due to the $SU(2)$ symmetry. We note that such $\Z2$ symmetry necessarily arises in spinful fermionic models with onsite disorder and $SU(2)$ symmetry~\cite{foot2} probed experimentally~\cite{1dcoupled,schreiber15}. We also demonstrated the resilience of the subspace dynamics in the presence of weak $\Z2$ symmetry breaking. An interesting possibility is to construct essentially arbitrary transport dynamics within an invariant subspace using engineered disorder, as well as to generalize these results to other models with more degrees of freedom per lattice site. 

T.I. is supported by a JQI postdoctoral fellowship, the Laboratory for Physical Sciences, and Microsoft.  M.\v Z. is supported by Grants No.~J1-7279 and No.~P1-0044 from the Slovenian Research Agency. We would like to thank KITP for its hospitality during the program ``The Dynamics of Quantum Information," where this work was initiated. This research was supported in part by the National Science Foundation under Grant No. NSF PHY-1748958.


\begin{thebibliography}{99}
\bibitem{foot3} Note that a global spin rotation suffices, as the doublon and holon states are $SU(2)_{\rm spin}$ singlets. In the spin language one would use $U=\exp{(-\ii \frac{\pi}{8}\sum_k \sx{k}\ty{k}-\sy{k}\tx{k}})$.
\bibitem{foot2} To have only $SU(2)$ symmetry without any global discrete one would require, e.g., bond-disordered models, which as far as localization is concerned are known to be different~\cite{vasseur15,romain16}.
\end{thebibliography}
\end{document}